# Accelerating the electronic-structure calculation of magnetic systems by equivariant neural networks

Yang Zhong[1,2], Binhua Zhang[1,2], Hongyu Yu[1,2], Xingao Gong[1,2], Hongjun Xiang[1,2*]

[1]Key Laboratory of Computational Physical Sciences (Ministry of Education), Institute of Computational Physical Sciences, State Key Laboratory of Surface Physics, and Department of Physics, Fudan University, Shanghai, 200433, China
[2]Shanghai Qi Zhi Institute, Shanghai, 200030, China
*E-mail: hxiang@fudan.edu.cn

## Abstract

Complex spin-spin interactions in magnets can often lead to magnetic superlattices with complex local magnetic arrangements, and many of the magnetic superlattices have been found to possess non-trivial topological electronic properties. Due to the huge size and complex magnetic moment arrangement of the magnetic superlattices, it is a great challenge to perform a direct DFT calculation on them. In this work, an equivariant deep learning framework is designed to accelerate the electronic calculation of magnetic systems by exploiting both the equivariant constraints of the magnetic Hamiltonian matrix and the physical rules of spin-spin interactions. This framework can bypass the costly self-consistent iterations and build a direct mapping from a magnetic configuration to the ab initio Hamiltonian matrix. After training on the magnets with random magnetic configurations, our model achieved high accuracy on the test structures outside the training set, such as spin spiral and non-collinear antiferromagnetic configurations. The trained model is also used to predict the energy bands of a skyrmion configuration of NiBrI containing thousands of atoms, showing the high efficiency of our model on large magnetic superlattices.

# Introduction

In recent years, some magnetic superlattices, such as skyrmion, spin spiral, and spin waves, have attracted much attention. Different from traditional ferromagnetism or antiferromagnetism, these magnetic superlattices often have non-trivial magnetoelectric coupling properties[1-7], such as the topological Hall effect, skyrmion Hall effect, etc., which are expected to be the material basis of new quantum devices with topological degrees of freedom. Various magnetic properties of materials come from the spin attribute of the electrons in different electronic states. However, the electronic structures of these nanoscale magnetic superlattices are hard to be calculated by traditional DFT methods. Recently, deep learning electronic structure methods provide a shortcut that can bypass the costly self-consistent iterations to obtain the ab initio electronic Hamiltonian of large systems[8-15]. Currently, several models for fitting the electronic Hamiltonian matrix of general non-magnetic materials have been reported[16-20]. However, the direct prediction of the Hamiltonian of magnetic systems becomes more challenging due to the existence of atomic magnetic moments. Recently Li et al. have reported a symmetry-constrained neural network model for predicting the electronic Hamiltonian of magnetic semiconductors[21]. Here we propose a deep learning model constrained by both the symmetry and physical rules, which can explicitly include all the important underlying spin interactions and fit the electron Hamiltonian of the magnetic system with fewer parameters.

For nonmagnetic systems, the electronic Hamiltonian can be regarded as a high-dimensional nonlinear function that depends only on the species $\{Z\}$ and positions $\{\vec{R}\}$ of the atoms in the system. However, for magnetic systems, the electronic Hamiltonian is also related to the magnitude and direction of the magnetic moment of each atom, which greatly increases the complexity of the problem. The magnetic Hamiltonian matrix needs to satisfy the SO(3) rotational equivariance and parity symmetry of atomic positions $\{\vec{R}\}$, the SU(2) rotational equivariance of spin bases $\{\vec{S}\}$, and the time reversal equivariance of magnetic moments $\{\vec{M}\}$ and spin bases $\{\vec{S}\}$. For different choices of the coordinate system in a crystal, the corresponding Hamiltonian matrix will be transformed between the different coordinate systems according to the equivariant constraints. For models that are not strictly equivariant, the Hamiltonian matrix may change unphysically when the system rotates or the coordinate system is changed. In addition, the Hamiltonian matrix of the magnetic system is still composed mainly of the contributions of non-magnetic interactions. Since spin-spin interactions are several orders of magnitude weaker than non-magnetic interactions, it is difficult to accurately capture the small effect of spin-spin interactions on the complete Hamiltonian of magnetic systems. Obviously, non-magnetic interactions and magnetic interactions should be treated differently with the constraints of physical rules to distinguish their differences in the contributions to the total Hamiltonian of magnetic systems.

In this work, we proposed an equivariant framework HamGNN++ for parameterizing magnetic Hamiltonian matrices. This framework can build a direct mapping from the crystal structures and magnetic configurations to the ab initio magnetic Hamiltonian matrices, thus accelerating the DFT calculation for large magnetic systems. In this framework, the magnetic Hamiltonian matrix is represented by an analytical parameterized formula that is constrained by the physical rules of spin-spin interactions and strictly satisfies both SU(2) and time-reversal equivariance. In the test, the trained HamGNN++ model successfully predicted the SOC magnetic Hamiltonian of spin-spiral, non-collinear antiferromagnetic, and skyrmion configurations, showing high accuracy and transferability. Our framework provides an efficient tool for the electronic structure calculation of large magnetic superlattices.

## Results

### *Theory*

For a non-magnetic system, the Hamiltonian matrix is uniquely determined by its atomic species $\{Z\}$ and coordinates $\{\vec{R}\}$. While the electronic Hamiltonian matrix for a magnetic system is also dependent on the atomic magnetic moments $\{\vec{M}\}$ and is calculated self-consistently by the non-collinear spin-constrained DFT calculations. The Hamiltonian matrix $H$ for magnetic systems can be generally written as the sum of the Hamiltonian matrix $H_0$ independent of the direction of the magnetic moments and $H_{mag}$ derived from the orientation constraints of the atomic magnetic moments. Both of the two parts are determined self-consistently, so the non-collinear spin-constrained DFT calculation requires more self-consistent iteration steps and is often difficult to converge for complex magnetic configurations. The direct prediction of the magnetic Hamiltonian matrix by deep neural networks can accelerate the non-collinear spin-constrained DFT calculations without convergence difficulties. The previous models for fitting the Hamiltonian of non-magnetic systems can not be directly extended to the magnetic systems[10, 16, 17], because the magnetic Hamiltonian $H$ also depends on the local magnetic moment $\{\vec{M}\}$ and satisfies the SU(2) and time reversal equivariance.

The magnetic Hamiltonian matrix elements obtained by the ab initio tight-binding methods can be regarded as a function of the atomic numbers $\{Z\}$, atomic coordinates $\{\vec{R}\}$, and the local atomic magnetic moments $\{\vec{M}\}$: $\hat{\mathbf{H}}^{s_i s_j}_{n_i l_i m_i, n_j l_j m_j}(\{Z\},\{\vec{R}\},\{\vec{M}\}) \equiv \langle \phi_{n_i l_i m_i} s_i | \hat{H} | \phi_{n_j l_j m_j} s_j \rangle$, where $\phi_{n_i l_i m_i}(\vec{r}-\vec{R}_i)$, $\phi_{n_i l_i m_i}(\vec{r}-\vec{R}_j)$ are the atomic orbital basis in real space, $s_i$, $s_j$ = ↑ or ↓. For a rotation inversion operation $gQ$, where $g \in \{I, E\}$, $Q \in SO(3)$, the magnetic Hamiltonian matrix elements satisfy the following SU(2) rotation equivariant constraint:

$$\langle gQ(\phi_{n_i l_i m_i} s_i) | \hat{\mathbf{H}} | gQ(\phi_{n_j l_j m_j} s_j) \rangle = \sigma_{p_i p_j}(g) \sum_{\mu_i=-l_i}^{l_i} \sum_{\mu_j=-l_j}^{l_j} \sum_{s'_i=-1/2}^{1/2} \sum_{s'_j=-1/2}^{1/2} \left( D^{(l_i)}_{m_i \mu_i}(Q) D^{(l_j)}_{m_j \mu_j}(Q) \right.$$

$$\left. D^{(1/2)*}_{s_i s'_i}(Q) D^{(1/2)}_{s_j s'_j}(Q) \langle \phi_{n_i l_i \mu_i} s'_i | \hat{\mathbf{H}} | \phi_{n_j l_j \mu_j} s'_j \rangle \right) \quad (1)$$

where $D(Q)$ is the Wigner $D$ matrix and $\sigma_p(g)$ is the scalar irreducible representation of the inversion operation, which is defined as follows

$$\sigma_p(g) = \begin{cases} 1, & g = E \\ p, & g = I \end{cases} \quad (2)$$

Now we consider the time-reversal equivariant constraint of the magnetic Hamiltonian matrix. The complete magnetic Hamiltonian consists of four sub-blocks, and the time-reversal equivariance of each block $\hat{\mathbf{H}}^{s_i s_j}(\{\vec{M}\}) \equiv \langle s_i | \hat{\mathbf{H}}(\{\vec{M}\}) | s_j \rangle$ is given by the following equation

$$\hat{\mathbf{H}}^{s_i s_j}(\{\vec{M}\}) = \langle T s_i | T \hat{\mathbf{H}}(\{\vec{M}\}) T^{-1} | T s_j \rangle^* = \langle T s_i | \hat{\mathbf{H}}(\{-\vec{M}\}) | T s_j \rangle^*. \quad (3)$$

According to $T|\uparrow\rangle = |\downarrow\rangle$, $T|\downarrow\rangle = -|\uparrow\rangle$, Eq. (3) can then be written as

$$\hat{\mathbf{H}}^{s_i s_j}(\{\vec{M}\}) = (-1)^{\delta_{s_i,s_j}} \left[ \hat{\mathbf{H}}^{(-s_i)(-s_j)}(\{-\vec{M}\}) \right]^*. \quad (4)$$

By combining the Hermitian symmetry of the Hamiltonian matrix, each block of the magnetic Hamiltonian matrix satisfies the following identity:

$$\hat{\mathbf{H}}^{s_i s_j}(\{\vec{M}\}) = \left( \hat{\mathbf{H}}^{s_i s_j}(\{\vec{M}\}) + (-1)^{\delta_{s_i,s_j}} \hat{\mathbf{H}}^{(-s_i)(-s_j)}(\{-\vec{M}\})^* \right) / 2 \quad (5)$$

Without loss of generality, each matrix subblock can be represented as the sum of the Hamiltonian matrices with odd and even time-reversal parity:

$$\hat{H}^{s_i s_j}\left(\{\vec{M}\}\right) = \tilde{H}_e^{s_i s_j}\left(\{\vec{M}\}\right) + \tilde{H}_o^{s_i s_j}\left(\{\vec{M}\}\right) \tag{6}$$

where $\tilde{H}_e^{s_i s_j}\left(\{-\vec{M}\}\right) = \tilde{H}_e^{s_i s_j}\left(\{\vec{M}\}\right)$, $\tilde{H}_o^{s_i s_j}\left(\{-\vec{M}\}\right) = -\tilde{H}_o^{s_i s_j}\left(\{\vec{M}\}\right)$.

Substituting Eq. (6) into Eq. (5), the general expression for the time-reversal equivariant magnetic Hamiltonian can be written as

$$\hat{H}^{s_i s_j}\left(\{\vec{M}\}\right) = \left(\tilde{H}_e^{s_i s_j}\left(\{\vec{M}\}\right) + (-1)^{\delta_{s_i,s_j}} \tilde{H}_e^{(-s_i)(-s_j)}\left(\{\vec{M}\}\right) + \right.$$
$$\left. \tilde{H}_o^{s_i s_j}\left(\{\vec{M}\}\right) + (-1)^{\delta_{s_i,s_j}+1} \tilde{H}_o^{(-s_i)(-s_j)}\left(\{\vec{M}\}\right)\right)^* \Big/ 2 \tag{7}$$

It can be seen from Eqs. (6) and (7) that fitting the magnetic Hamiltonian directly from the symmetry constraint requires eight sub-blocks with different time-reversal parity and spin components. This method relies on at least eight times the parameters and computations than in the non-magnetic case. In this work, we combine both the equivariant constraints and physical rules to fit the magnetic Hamiltonian matrix. Compared with pure data-driven learning, our framework imposes physical information constraints in the model and can achieve more generalization ability with fewer network parameters.

The complex non-collinear magnetic configurations are usually formed under the influence of multiple types of spin-spin interactions, such as the Heisenberg exchange, the Dzyaloshinskii–Moriya interaction[22], and the Kitaev-type exchange[23, 24]. The spin-spin interaction energy can be quantitatively calculated by the effective spin Hamiltonian method, which is widely used to explain and simulate various complex magnetic configurations. By explicitly embedding the prior knowledge of the spin-spin interactions in the model, we can reduce the redundant parameters of the network. Starting from the Heisenberg model, we derived an analytical parameterized formula for the magnetic Hamiltonian matrix, which depends on about 1/8 of the parameters in Eq. (7) and satisfies the fundamental equivariant constraints of the magnetic Hamiltonian matrix.

The classical Heisenberg model can be written in a general matrix form containing all possible second-order interactions[25]:

$$E_{spin} = \sum_{\langle i',j' \rangle} \sum_{\alpha\beta} J_{i'j'}^{\alpha\beta} \left( M_{i'}^{\alpha}[\rho] \cdot M_{j'}^{\beta}[\rho] \right) + \sum_{k'} \sum_{\alpha\beta} A_{k'}^{\alpha\beta} \left( M_{k'}^{\alpha}[\rho] \cdot M_{k'}^{\beta}[\rho] \right), \qquad (8)$$

where the 3×3 tensors $J_{i'j'}^{\alpha\beta}$ and $A_{k'}^{\alpha\beta}$ are called the $J$ matrix and single-ion anisotropy (SIA) matrix, respectively. $M_{i'}^{\alpha}[\rho]$ is the component of the ionic magnetic moment $\vec{M}_{i'}[\rho] = Tr\left[ \rho \hat{W}_{i'} \hat{\vec{\sigma}} \right]$, which is a functional of the electron density $\rho$. The magnetization of the system is partitioned into the local magnetic moments on each atom by a pre-defined weight operator $\hat{W}_{i'}$, which is commonly defined as $\hat{W}_{i'} = \int d\vec{r}\, f\left( r_{cut} - \|\vec{r} - \vec{\tau}_{i'}\| \right) |\vec{r}\rangle\langle\vec{r}|$, where $f$ is a radial cutoff function centered on the atom $i'$. The matrix element of the weight operator $\hat{W}_{i'}$ is given by[26, 27]

$$\left( \hat{W}_{i'} \right)_{n_i l_i m_i, n_j l_j m_j} = \begin{cases} w_{n_i l_i m_i, n_j l_j m_j}, & i, j = i' \\ \dfrac{1}{2} w_{n_i l_i m_i, n_j l_j m_j}, & i \text{ or } j = i' \\ 0, & i, j \neq i' \end{cases} \qquad (9)$$

The matrix $w_{n_i l_i m_i, n_j l_j m_j}$ varies with the choice of the cutoff function $f$ and is regarded as a learnable parameter in our model. The magnetic part of the parameterized Hamiltonian for the magnetic systems equals the variational derivative $\dfrac{\delta E_{spin}}{\delta \rho}$:

$$\tilde{H}^{mag}_{n_i l_i m_i s_i, n_j l_j m_j s_j} = \sum_{\langle i',j' \rangle} \sum_{\alpha\beta} J^{\alpha\beta}_{i'j'} \left( M^{\alpha}_{i'} \left(W_{j'}\right)_{n_i l_i m_i, n_j l_j m_j} \sigma_{\beta, s_i s_j} + M^{\beta}_{j'} \left(W_{i'}\right)_{n_i l_i m_i, n_j l_j m_j} \sigma_{\alpha, s_i s_j} \right) +$$

$$\sum_{k'} \sum_{\alpha\beta} A^{\alpha\beta}_{k'} \left(W_{k'}\right)_{n_i l_i m_i, n_j l_j m_j} \left( M^{\alpha}_{k'} \sigma_{\beta, s_i s_j} + M^{\beta}_{k'} \sigma_{\alpha, s_i s_j} \right). \quad (10)$$

The $J^{\alpha\beta}_{i'j'}$ and $A^{\alpha\beta}_{k'}$ cartesian tensors are learnable and can be mapped equivariantly from the features of the edges and nodes respectively. Since the spin magnetic moment is odd under time-reversal operation and even under spatial inversion operation, the cartesian tensors $J^{\alpha\beta}_{i'j'}$ and $A^{\alpha\beta}_{k'}$ should be even under time-reversal and spatial inversion operations so that the parameterized Hamiltonian matrix constructed by Eq. (10) satisfies the time-reversal equivariance and the parity symmetry. In addition, Eq. (10) still satisfies all the equivariance when only the ionic magnetic moments are rotated.

According to the expression of the parameterized magnetic Hamiltonian matrix proposed in this work, we designed the HamGNN++ network to predict the magnetic Hamiltonian matrix based on HamGNN[20]. HamGNN++ builds an equivariant mapping from the crystal graph containing the atomic numbers $\{Z_i\}$, the atomic coordinates $\{\vec{R}_i\}$, and the atomic magnetic moments $\{\vec{M}_i\}$ to the magnetic Hamiltonian. The one-hot embeddings of the atomic numbers $\{Z_i\}$ and the magnitudes of atomic magnetic moments $\{|\vec{M}_i|\}$ are combined and passed through an MLP layer to form the initial atomic features $V_i^{(0)}$. The initial atomic features $V_i^{(0)}$ contain only the invariant features of $l = 0$, and higher-order equivariant features with $l > 0$ are initialized to 0. The distance $|\vec{r}_{ij}|$ between the central atom $i$ and the neighbor atom $j$ is expanded into a set of Bessel function bases $B(|\vec{r}_{ij}|)$. The direction between $i$ and $j$ is embedded as a set of real spherical harmonic basis $Y(\vec{r}_{ij})$. The initial atomic features are refined by T orbital convolution layers, which update the features of the central atoms $i$ by aggregating the equivariant messages of neighbor atoms $j$. The equivariant messages of neighbor atom $j$ are obtained by the tensor product between the features of atom $j$ and the spherical harmonic basis of the inter-atomic orientation $\hat{r}_{ij}$. The final updated atomic features $V_i^{(T)}$ are passed through the on-site bocks to obtain spin-less on-site Hamiltonian $H_{ii}^{(0)}$ as well as single-ion anisotropy parameters $A_{ii}$. The central atom features $V_i^{(T)}$ and the neighbor atom features $V_j^{(T)}$ are used as the input of the pair interaction block to construct the pair features $P_{ij}$ of each edge, which are mapped to spin-less off-site Hamiltonian $H_{ij}^{(0)}$ and magnetic coupling parameters $J_{ij}$ through the off-site blocks.

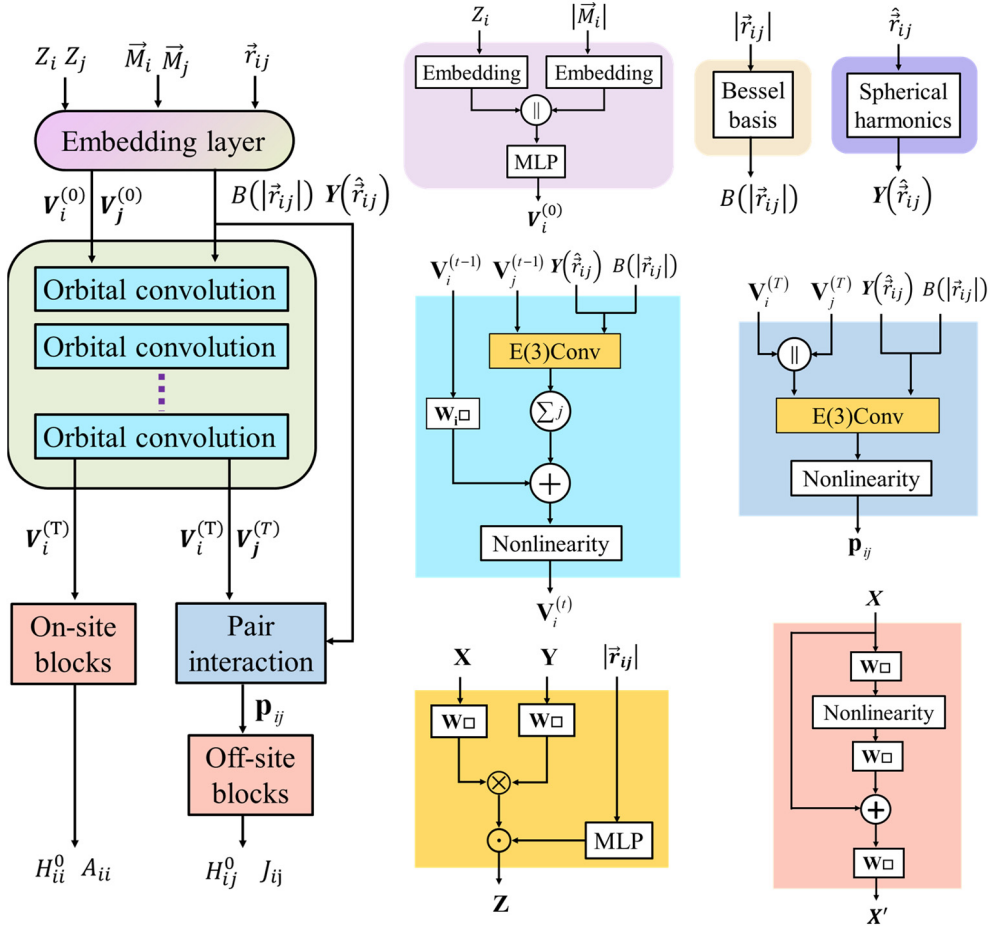

**Figure** 1. The network architecture of HamGNN++. The atomic number and the magnitude of the atomic magnetic moments are encoded as one-hot vectors respectively and then merged into the initial node features $V_i^{(0)}$ by an MLP. "∥" denotes the concatenation operator of the vectors. Bessel basis and spherical harmonic functions are used to expand the distances and orientations between adjacent atoms respectively. The initial node features are updated through T orbital convolution layers, which use E(3) equivariant convolution to update the node features of the central atoms by aggregating the information of neighbor atoms. ⊗ denotes the tensor product, and ⊙ denotes the element-wise multiplication. The refined node features $V_i^{(T)}$ can output the on-site parameters such as $H_{ii}^{(0)}$ and $A_{ii}$ through the on-site blocks. The pair interaction layer builds the pair features $P_{ij}$ of each edge from the features of the central atoms $i$ and the neighbor atoms $j$. The off-site parameters ($H_{ij}^{(0)}$ and $J_{ij}$) are mapped from the pair features $P_{ij}$ through the off-site block.

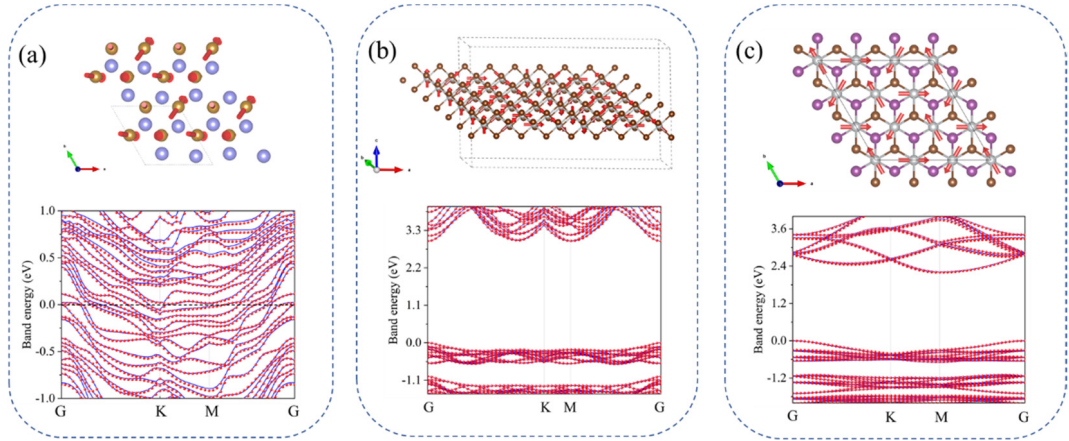

**Figure 2**. Test results of HamGNN++ on the magnetic configurations of monolayers of Fe/Ir(111), NiBr$_2$, and NiBrI. (a) Comparison of the band structures of HamGNN++ prediction (solid line) and DFT calculation (scatter points) for a 2×2×1 Fe/Ir (111) monolayer with random magnetic moment orientations of Fe. (b) Comparison of the band structures of HamGNN++ prediction(solid line) and DFT calculation (scatter points) for a NiBr$_2$ monolayer with 8×1×1 spiral magnetic configuration of Ni. (c) Comparison of the band structures of HamGNN++ prediction (solid line) and DFT calculation (scatter points) for a 3×3×1 NiBrI monolayer with random magnetic moment orientations of Ni.

## Tests

HamGNN++ was used to fit the Hamiltonian of the magnetic configurations of Fe/Ir(111), NiBr$_2$, and NiBrI crystals. Fe/Ir(111) is metallic, while NiBr$_2$ and NiBrI are magnetic semiconductors. Neel-type skyrmions were first found in monolayer Fe/Ir (111) film[3], whose SOC effect and DM interaction are very strong. Fitting the SOC Hamiltonian matrix of monolayer Fe/Ir (111) film is a challenging task. The non-collinear spin-constrained DFT calculations on the Fe/Ir(111) metallic film with some complex magnetic configurations are hard to converge, while HamGNN++ can directly map the atomic coordinates $\{\vec{R}_i\}$ and atomic magnetic moments $\{\vec{M}_i\}$ to the magnetic Hamiltonian matrices without convergence difficulties. 1000 random crystal structures of Fe/Ir(111) supercells were obtained by applying random displacements of up to 0.1 Å to the atoms of 2×2×1 Fe/Ir(111) monolayers. We used OpenMX[28, 29] to perform the non-collinear spin-constrained DFT calculations on the Fe/Ir(111) supercells with the magnetic moments of the Fe atoms constrained in random directions and divided the calculated Hamiltonian matrices into the training, validation, and test sets in a ratio of 0.8: 0.1: 0.1. The mean absolute error (MAE) of the predicted Hamiltonian matrix on the test set is as low as sub-meV. The calculated and predicted band structure of a Fe/Ir(111) supercell with random magnetic configuration in the test set is in good agreement, as shown in Fig. 2(a).

The random magnetic configurations of 1000 NiBr$_2$ 3×3×1 supercells and 1000 NiBrI 2×2×1 supercells were constructed in the same way as that in Fe/Ir(111) supercells, and the Hamiltonian matrices calculated by spin-constrained non-collinear DFT were divided into the training, validation, and test sets in a ratio of 0.8: 0.1: 0.1. We constructed the spin-spiral configuration of NiBr$_2$ and the non-collinear antiferromagnetic configuration of NiBrI respectively and used the trained HamGNN++ model to predict the energy bands of these two test structures, as shown in Fig. 2(b) and (c). The predicted bands accurately reproduce the energy bands calculated by DFT, showing that the HamGNN++ model has good generalization performance.

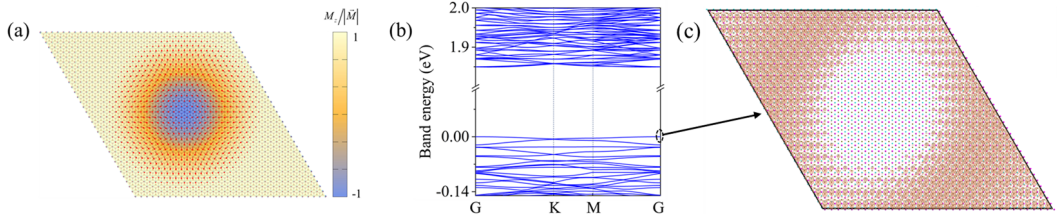

**Figure 3**. Prediction results of the antiskyrmion magnetic configuration of NiBrI by HamGNN++. (a) Crystal structure of monolayer NiBrI with antiskyrmion magnetic configuration. The crystal structure contains 3468 atoms, forming an antiskyrmion configuration with a radius of about 10 Å. (b) The predicted band structure of NiBrI with antiskyrmion magnetic configuration. (c) The charge density of the VBM wave function.

The magnetic superlattices, represented by the skyrmions[30-32], have many non-trivial electronic properties, but due to their huge sizes, it is difficult to calculate their electronic structures directly by DFT methods. Using the neural network method proposed in this work, we can make a direct prediction with the DFT level for the electronic structures of large magnetic superstructures. As shown in Fig. 3(a), we constructed an antiskyrmion magnetic configuration in a 34×34×1 single-layer NiBrI supercell (with a total of 3468 atoms) and calculated its band structure with the Hamiltonian predicted by HamGNN++, as shown in Fig. 3(b). The predicted bandgap of this antiskyrmion magnetic configuration is 1.85 eV. The valence band maximum (VBM) appears at point G and the VBM wave function is highly localized in the region between skyrmion patterns, as shown in Fig. 3(c). In addition, The Dirac cones at point $K$ shows that this magnetic configuration has non-trivial topological properties.

## Discussion

We proposed a parameterization scheme for the magnetic Hamiltonian based on the effective model of spin-spin interactions, which significantly reduces the number of required parameters for prediction and satisfies the $SU(2) \otimes \{I, T\}$ equivariance. Based on this parameterization scheme, we designed an equivariant HamGNN++ deep neural network, which can build a mapping from the magnetic configuration to the electron Hamiltonian matrix and greatly reduce the computational complexity. The network was trained on the random magnetic configurations of metallic Fe/Ir(111) thin film and two monolayer magnetic semiconductors ($NiBr_2$ and NiBrI). The electronic structures of the magnetic configurations outside the training set were successfully predicted by the trained model with only a small error compared with DFT. We also used the trained model to calculate the band structure of the Skyrmion configuration of NiBrI to demonstrate the high efficiency of the model for calculating magnetic superlattices. The proposed parameterization scheme for the magnetic Hamiltonian provides an efficient and accurate tool for studying the electronic structure of magnetic superlattices.